\documentclass{emulateapj}

\usepackage{apjfonts}

\newcommand{\msun}{{M_{\odot}}}
\newcommand{\Mpc}{\mathrm{Mpc}}
\newcommand{\dmass}{\delta_{\mathrm{mass}}}
\newcommand{\dhalo}{\delta_{\mathrm{halo}}}
\newcommand{\hiMpc}{h^{-1}\Mpc}
\newcommand{\hiMsun}{h^{-1}\msun}

\shorttitle{Proto-clusters in $\Lambda$CDM Universe}
\shortauthors{Suwa et al.}

\begin{document}

\title{Proto-clusters in the $\Lambda$CDM Universe}
\author{Tamon Suwa\altaffilmark{1}, Asao Habe\altaffilmark{2}, and Kohji
Yoshikawa\altaffilmark{3}}

\altaffiltext{1}{Center for Computational Sciences, University of
Tsukuba, Tsukuba 305-8577, Japan; tamon@ccs.tsukuba.ac.jp}
\altaffiltext{2}{Division of Physics, Graduate School of Science, Hokkaido
University,Sapporo 060-0810, Japan; habe@astro1.sci.hokudai.ac.jp}
\altaffiltext{3}{Department of Physics, School of Science, The
University of Tokyo, 7-3-1 Hongo, Bunkyo-ku, Tokyo 113-0033, Japan;
kohji@utap.phys.s.u-tokyo.ac.jp}

\begin{abstract}
We compare the highly clustered populations of very high redshift
galaxies with proto-clusters identified numerically in a standard
$\Lambda$CDM universe  ($\Omega_0=0.3,\, \lambda_0=0.7$) simulation.
We evolve $256^3$ dark matter particles in a comoving box of side
 $150h^{-1}\Mpc$.
By the present day there are 63 cluster sized objects of mass in excess
 of $10^{14}h^{-1}\msun$ in this box.
We trace these clusters back to higher redshift finding that their
 progenitors at $z=4$--$5$ are extended regions of typically 20--40 Mpc
 (comoving) in size, with dark halos of mass in excess of
 $10^{12}h^{-1}\msun$ and are overdense by typically 1.3--13 times the
 cosmological mean density.
 Comparison with the observation of Ly$\alpha$ emitting (LAEs) galaxies
 at $z=4.86$ and at $z=4.1$
 indicates that the observed excess clustering is consistent with that
 expected for a proto-cluster region if LAEs typically correspond to
 massive dark halos of more than $10^{12}h^{-1}\msun$.
We give a brief discussion on the relation between high redshift
 concentration of massive dark halos and present day rich clusters of
 galaxies.
\end{abstract}

\section{Introduction}
The formation and evolution of structures in the universe, such as
galaxy clusters and large-scale structures, are part of the most
important issue in astrophysics.
Since clustering properties of galaxies in the distant universe give us
great clues to this issue, many observations have been done in order to
detect (proto) clusters of galaxies at high redshifts \citep{Steidel98,
Steidel00,  Campos99, Pentericci00, Rhoads00, Ouchi01, Ouchi03, Miley04}.
These deep surveys of galaxies, which have significantly advanced our 
understanding of the properties and distribution of galaxies at high 
redshifts, are equally important in constraining the underlying 
structure formation scenarios.

\citet{Venemans02} showed a region around a radio galaxy in which number
density of Ly$\alpha$ emitters (LAEs) is much higher than the mean of
the universe suggested by \citet{Rhoads00} at $z=4.1$.
The region has size of 2.7 $\times$ 1.8 Mpc (physical) and mass of $\sim
10^{15}\msun$.
More recently, \citet{Shimasaku03} found that LAEs are clustered in an
elongated region with a size of 20 $\times$ 50 Mpc (comoving) at
$z=4.86$, which is comparable to the size of present-day large-scale
structures.
In this elongated region, there is a circular region of high surface
density of LAEs with 12 Mpc radius, which may be a progenitor of a galaxy
cluster.
They also estimated the bias parameter of the LAEs in the range of 
$b\sim 3$--$16$  for spatially flat low-density cosmological model
($\Lambda$CDM model).
The elongated distribution of the LAEs in this region is also proposed
to be a part of large-scale structures, although observation by
\citet{Shimasaku04} of the same sky area which is closer to us by 39Mpc
(comoving) does not show large structure.
In addition, \citet{Hayashino04} reported that LAEs distribution at
$z=3.1$ in their observed area shows broad large scale structure.
They claim that this structure cannot be explained in the context of the
standard CDM model.
\citet{Ouchi05} observed LAEs at $z=5.7$ and discovered filamentary
structures and voids.

Although these regions which have much high number densities of LAEs are
claimed to be proto-clusters, there is not yet enough detailed study on
these possibilities.
Therefore, we investigate formation and evolution of galaxy clusters in
early universe using cosmological simulation.
In this letter we report some properties of simulated proto-clusters at
$z=5$ and compare our numerical results with observations by
\citet{Shimasaku03} and \citet{Venemans02}.

\section{Method}
\label{sec:method}
\subsection{Numerical Method}
We perform a cosmological N-body simulation with
Particle-Particle-Particle-Mesh ($\mathrm{P^3M}$) algorithm\citep{HE81}.
Our simulation code is the same one which was used in
\citet*{YJS00}.
We use the following values as parameters of our simulation:
the Hubble constant in units of
$100 \mathrm{km\ s^{-1}\ Mpc^{-1}}$, $h=0.7$, the density parameter,
$\Omega_0 = 0.3$, the baryon density parameter, $\Omega_b = 
0.015h^{-2}$, the root mean square density fluctuation amplitude on a
scale $8h^{-1}\mathrm{Mpc}$, $\sigma_8=1.0$, the power-law index of
the primordial density fluctuation, $n=1.0$, and the cosmological constant
parameter, $\lambda_0=0.7$.
We employ $N_{\mathrm{DM}}=256^3$ dark matter particles and
the mass of each one is $2.15 \times 10^{10}\msun$.
The size of the comoving simulation box, $L_{\mathrm{box}}$, is
$150 h^{-1} \mathrm{Mpc}$, and the box is on the periodic boundary
condition.
This size of the box is large enough to realize a sufficient number
of clusters and large-scale structures at $z=0$.
We use the spline (S2) softening function for gravitational softening,
and the softening length, $\epsilon_{\mathrm{grav}}$, is set to be
$L_{\mathrm{box}}/(10N_{\mathrm{DM}}^{1/3})$ ($\sim 60h^{-1}
\mathrm{kpc}$ in the comoving scale).

\subsection{Identification of Dark Halos and Proto-clusters}
We identify dark halos in a manner similar to that of \citet{Suwa03}.
A brief explanation of the identification is as follows:
First, we define densities of dark matter particles using an
interpolation technique in the same way of the Smoothed Particle
Hydrodynamics method \citep{Monaghan92}, and pick up 
particles whose densities are more than the virial density.
Next, we perform the hierarchical friends-of-friends (HFoF) method
\citep{Klypin99} for the selected dense particles.
The maximum linking length $l_{\mathrm{max}}$ in HFoF method is defined
as $0.2\bar{l}$, where $\bar{l}$ is the mean inter-particle distance for
all (not only dense) particles.
Then we draw a sphere of which the center is on the densest particle of
the group and seek the radius in which the mean density of total matter
is equal to the virial density.
We regard the set of particles in the sphere as a virialized object.
If a centroid of a sphere exists in another sphere, only the set which
belongs to the more massive sphere is used.
We regard dark halos which have mass of more than 
$10^{14} h^{-1}\msun$ at $z=0$ as galaxy clusters and we analyze
properties of their progenitors to compare with the observations of
proto-cluster regions \citep{Venemans02, Shimasaku03}.

We seek proto-clusters in high redshift universe as follows: 
First, we pick up all particles in each cluster at $z=0$.
Next, we trace those particles back to high-z epoch, e.g.~$z=5$.
Finally, we set a minimum cubic region which covers all of the particles.
We call the region as ``proto-cluster region'' for the cluster.
This region contains the proto-cluster of the present cluster.

In order to investigate galaxy distribution in the proto-cluster
regions, we identify galaxy-size dark halos ($M > 10^{12}h^{-1}\msun$) 
in the regions at $z=5$.
Dark halos of this mass scale are suggested to contain LAEs and
LBGs from their spatial distribution arguments \citep{Hamana04}.

\subsection{Overdensity}
In proto-cluster regions, it is expected that dark halos which
correspond to galaxies concentrate much denser than the background.
Therefore the excess of number density of dark halos, $\dhalo$, should be
useful quantities to compare the result of the simulation to
observations:
\begin{equation}
 \dhalo = \frac{\bar{n}_{\mathrm{halo}}}{n_{\mathrm{halo,BG}}}-1,
\end{equation}
where $\bar{n}_{\mathrm{halo}}$ is the halo number density averaged in
the suitable scale (e.g.~25 Mpc) and $n_{\mathrm{halo,BG}}$ is the
background halo number density.

The mass overdensity, $\dmass$, is defined in a similar way:
\begin{equation}
 \dmass = \frac{\rho}{\rho_{\mathrm{BG}}}-1,
\end{equation}
where $\rho$ is the density of the region under consideration and 
$\rho_{\mathrm{BG}}$ is that of the background.
The combination of the halo overdensity and the mass overdensity give us
important information of mass concentration, so-called bias information,
which has been investigated in analytical and numerical works
\citep[e.g.][]{Kaiser84,TS00,Yoshikawa01}.

\section{Results}
\begin{figure}
\plotone{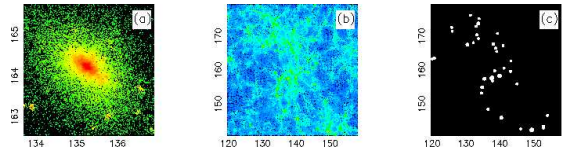}
\caption{Distributions of dark matter and dark halos in an example
 cluster and the corresponding proto-cluster region in the simulation box.
 Panel (a) and (b) show the distribution of dark matter at $z=0$ and $z=5$,
 respectively.
 Panel (c) shows the distribution of dark halos which have mass of more
 than $10^{12}h^{-1}\msun$ in the proto-cluster region shown in
 the panel (b).
$x$- and $y$-axes are comoving position coordinate in the simulation.}
\label{fig:plot_ex}
\end{figure}

We find 63 galaxy clusters with  $M >10^{14}h^{-1}\msun$  in the
simulation box at the present epoch ($z=0$).
We seek their progenitors, i.e.~proto-clusters, in high redshift universe and
study their properties.
We show an example of present galaxy clusters and proto-cluster regions in our
simulation box at $z=5$ in Figure \ref{fig:plot_ex}.

In our numerical results, the size of the proto-cluster regions is in
the range of 20--40Mpc in the comoving scale.
The extension is similar to that of the LAEs dense region observed by
\citet{Shimasaku03}.
Several dark halos with $M > 10^{12} h^{-1}\msun$ already exist in
proto-cluster regions at $z=5$ as shown in Figure \ref{fig:plot_ex}(c).
We obtain bulk velocity of each dark halo in the proto-cluster
regions and found that dispersion of those are typically $\sim
200 \mathrm{km\ s^{-1}}$.

\begin{figure}
\includegraphics[height=0.4\textwidth,angle=270]{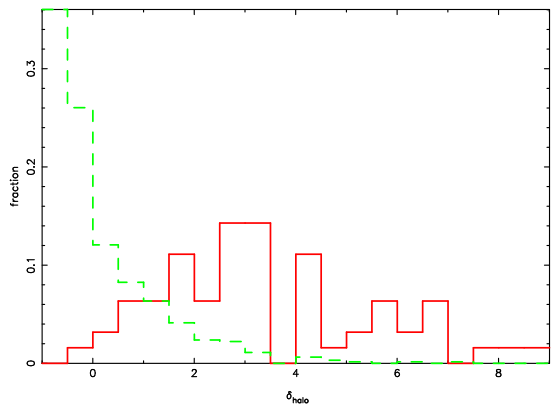}
\caption{Histogram of halo number overdensities, $\dhalo$, of
 the proto-cluster regions (red solid line) and the randomly selected region
 (green dashed line) at $z=5$.}
\label{fig:histogram_dhalo}
\end{figure}

We calculate $\dhalo$ and $\dmass$ for each proto-cluster regions at $z=5$,
by assuming a smoothing scale of 25 Mpc (comoving) that is a typical
proto-cluster size in our numerical results.
In order to show that the values of these indicators have
significant excess than the mean value of the universe, we randomly
select $(25 \Mpc)^3$ regions in the simulation box and calculate
$\dhalo$ and $\dmass$ for these regions.
The number of the randomly selected regions is 630, which is 10 times
more numerous than the proto-cluster regions.
We show the histograms of $\dhalo$ in Figure \ref{fig:histogram_dhalo},
where the red solid line is for the proto-cluster regions and the
green dashed line is for the randomly selected regions.
The red solid line is significantly different from the
green dashed line.
Figure \ref{fig:histogram_dhalo} shows that the typical value of
$\dhalo$ is $\sim 3$ for the proto-cluster regions, and its variance is
very large.
On the other hand, $\dhalo$ in most of the randomly selected regions are
almost $-1$, where this value means that there is no massive dark halo in the
regions.

\begin{figure}
\includegraphics[height=0.4\textwidth,angle=270]{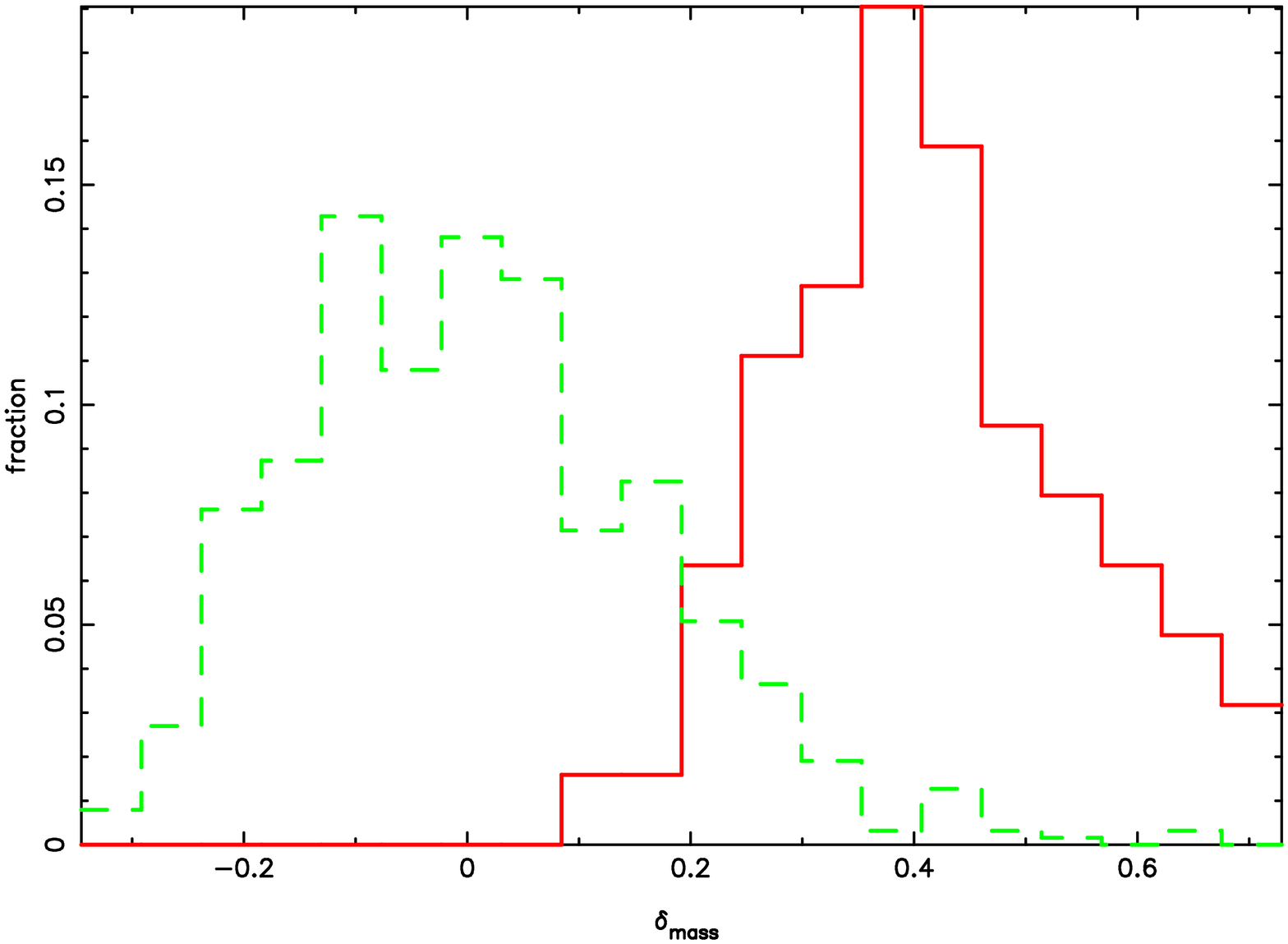}
\caption{Same as Figure \ref{fig:histogram_dhalo}, but for $\dmass$.}
\label{fig:histogram_dmass}
\end{figure}

The mass overdensities, $\delta_{\mathrm{mass}}$, for the proto-cluster
regions are clearly different from the randomly selected regions
as shown in Figure \ref{fig:histogram_dmass}.
$\delta_{\mathrm{mass}}$ of the proto-cluster regions are in the range
of 0.2--0.4, while those of the randomly selected regions range from 
$-0.2$ to $0.2$.
The typical value of $\dmass$ of the proto-cluster regions is $\sim 0.4$
and the variance of $\dmass$ is much smaller than that of $\dhalo$.

\begin{figure}
\includegraphics[height=0.4\textwidth,angle=270]{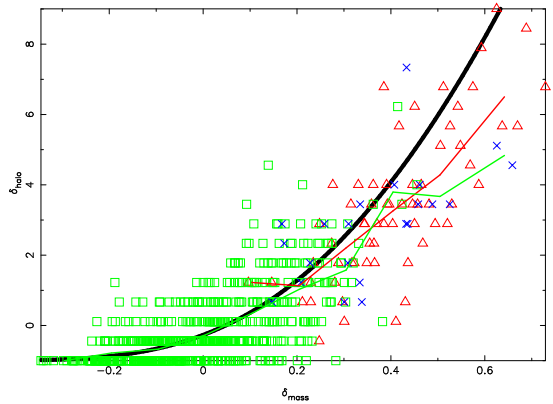}
\caption{Correlation map of $\dmass$ and $\dhalo$.
 Red triangles, blue crosses, and green squares indicate the
 proto-cluster regions, the randomly selected regions which overlap more
 than 50\% with some proto-cluster region, and other randomly selected
 regions, respectively.
The analytical relation is indicated by a thick black line.
 Green and red lines are correspond to average value
of simulated regions randomly selected (both field and overlapped) and proto-clusters, respectively.
}
\label{fig:scat_dm_dh}
\end{figure}

In order to show the bias of dark matter halo distribution in the
proto-cluster regions, we plot $\dmass$ and $\dhalo$ for the
proto-cluster regions and the randomly selected regions in Figure
\ref{fig:scat_dm_dh}.
Red triangles, blue crosses, and green squares indicate the
proto-cluster regions, the randomly selected regions which overlap more
than 50\% with some proto-cluster region, and other randomly selected
regions, respectively.
Figure \ref{fig:scat_dm_dh} shows correlation between the values of
$\dhalo$ and $\dmass$ for $\dmass > 0$, although the dispersion is very
large.
It is clear that the regions with large $\dmass$ have large $\dhalo$.
This relation corresponds to the bias of dark matter halo distribution in
the proto-cluster regions.

We compare the $\dmass$-$\dhalo$ relation in our numerical results with the
analytical result given by the natural bias theory \citep{MW96}
in Figure \ref{fig:scat_dm_dh}.
In Figure \ref{fig:scat_dm_dh}, we show the analytical relation by a thick
black line, where we use the Press-Schchter mass function formula and the
relation of linear extrapolation to nonlinear evolution for
density perturbation \citep*{CPT92}.
Green and red lines in Figure \ref{fig:scat_dm_dh} indicate average value
of the randomly selected regions and the proto-clusters, respectively.
The relation between the $\dmass$-$\dhalo$ given by the natural
bias theory agrees well with our numerical results as shown in Figure
\ref{fig:scat_dm_dh}, except for gradual deviation $\dmass >0.4$.
This deviation may be explained by the effect of limitation of our
simulation box size.
Our simulation box size is not enough to consider large-scale (100$\hiMpc$) component of density fluctuation and this limitation may cause
underestimation of mass function for $10^{12}\msun$ dark halos at
$z=5$ \citep{BR05}.

\begin{figure}
\includegraphics[height=0.4\textwidth,angle=270]{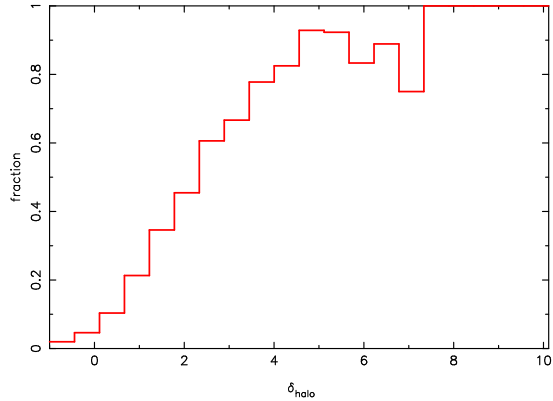}
\caption{Fraction of randomly selected cubic regions which contain a
 galaxy cluster at $z=0$ as a functions of $\dhalo$ at $z=5$.
 The size of the cubic regions is 25 Mpc (comoving).
}
\label{fig:dh_z5_Mmax_z0}
\end{figure}

It is very important to show a critical value of $\dhalo$,
$\delta_{\mathrm{halo,c}}$, by which we can select a proto-cluster
region.
In order to search $\delta_{\mathrm{halo,c}}$, we  compare values of
halo overdensity, $\dhalo$, in the 
randomly selected regions at $z=5$ and masses of the largest dark halos
in the regions at the present epoch.
Figure \ref{fig:dh_z5_Mmax_z0} shows a fraction of the regions which
contain galaxy clusters with $M >10^{14}\hiMsun$ at $z=0$ as a function of
$\dhalo$ at $z=5$.
It is clear in Figure \ref{fig:dh_z5_Mmax_z0} that most regions with
$\dhalo \ge 3$ at $z=5$ will evolve rich clusters with mass of
more than $10^{14}\hiMsun$ at $z=0$.

\section{Discussion}
We have investigated proto-cluster regions and the difference between
those regions and background field regions using N-body cosmological
code.
In order to investigate proto-clusters, we identify 63 galaxy
clusters at in our cosmological simulation $z=0$, and trace dark matter particles which belong to
those, back to a high redshift epoch.
We study properties of the proto-cluster regions by analyzing the mass
overdensity, $\dmass$, and the halo overdensity, $\dhalo$, of which
definitions are given  in \S\ref{sec:method}.
The results at $z=5$ are shown in Figures \ref{fig:histogram_dhalo},
\ref{fig:histogram_dmass}, and \ref{fig:scat_dm_dh}.
As expected from the bias theory \citep[e.g.][]{TS00}, $\dhalo$ shows
excess over $\dmass$ and variance of $\dhalo$ is much larger than that
of $\dmass$.

In order to show difference between the proto-cluster regions and the
field, we calculate $\dmass$ and $\dhalo$ for the randomly selected
regions with the same size of the typical proto-cluster region at $z=5$ 
(25 Mpc in the comoving scale).
We select 630 regions (10 times greater than the proto-cluster regions) for
the randomly selected regions.
Clustering properties of the dark halos in these regions are significantly
different from those in the proto-cluster regions.
The halo overdensities for most of the randomly selected regions are much smaller
than those for the proto-cluster regions as shown in Figure
\ref{fig:histogram_dhalo}.

It is very interesting to study a critical value of $\dhalo$
by which we can select a region as the proto-cluster regions.
From Figure \ref{fig:dh_z5_Mmax_z0}, we have found that more than 80\%
of the regions with $\dhalo \ge 3$ at $z=5$ contain rich clusters ($M \ge
10^{14}\hiMsun$) at the present epoch.
Thus, we conclude that $\dhalo \ge 3$ is a good criterion to distinguish
proto-cluster regions from field regions at $z=5$.
This criterion is very useful to judge whether observed galaxies excess
regions are actual proto-clusters or not.

It should be pointed out that there are a few regions which do not
contain clusters at $z=0$ but have large $\dhalo$ values (3--7) at $z=5$,
although a fraction of those regions are very small.
We discuss that the physical origin of the large $\dhalo$ is 
non-linearity and stochasticity of bias parameters 
proposed by \citet{TS00}.
They studied variance of biasing parameters, and conclude that the
variance increases strongly with redshift.
They also conclude that stochasticity of the biasing is generated by the
scatter in the halo mass distribution at higher redshift.
\citet{Yoshikawa01} confirmed these results for $0 < z< 3$ by large
$\mathrm{P^3M}$ simulation.
We conclude that our results for $z=3$ are consistent with their
results and the trend of evolution of bias variance is still true for 
$3 < z < 5$.

We compare our results with the observation of \citet{Shimasaku03}.
Properties of the LAEs concentrated region reported by them are as
follows:
(1) Diameter of the region is 25 Mpc (comoving unit),
(2) Projected overdensity of LAEs $\delta_{\Sigma} \sim 2$,
(3) Bias parameter is estimated 3--16 for $\Lambda$CDM model, and the
best-fit value is $b\sim 6$, i.e.~$\dmass\sim 0.3$.
(4) The number of LAEs in the region is about 20.
Properties from (1) to (3) are consistent with our numerical results of
proto-cluster regions, if LAEs correspond to dark halos with mass more than
$10^{12}h^{-1}\msun$, which is suggested by \citet{Hamana04} using
correlation function on small-scales.
The typical number of dark halos in the simulated proto-clusters is 5--10,
which is about half of that of observed LAEs.
One explanation of this discrepancy is due to possibility that some dark halos have more
than one LAEs.
We speculate that some pairs of LAEs in Fig.~3 of \citet{Shimasaku03}
are included in the same dark halo.
It is also possible that some of observed sample of LAEs in the region
are low redshift interlopers.
\citet{Shimasaku03} estimate the contamination of their sample to be
about 20\%.

\citet{Venemans02} also reported properties of a LAEs rich region.
The size of the region is $\sim$ 14 $\times$ 10 Mpc (comoving unit).
They estimate the region is overdense in LAEs by a factor of 15 compared
with the blank field \citep{Rhoads00}.
The overdensity is much larger than that of proto-clusters at $z=4$ in
our simulation and the size of their observed region is smaller than the
typical size of the proto-clusters in our simulation.
Therefore, we suggest that the region observed by them is not a whole
proto-cluster region, but the central region of the proto-cluster
because of small size of the observed region and their high value of the
overdensity.

\acknowledgements
We would like to thank Prof.~Masayuki Fujimoto and Dr.~Takayuki Saitoh
for helpful advice, insightful discussions and encouragement.
We are also grateful to Prof.~Tom Broadhurst for useful comments.
Numerical computations in this work were carried out on SGI Onyx300 in the
Hokkaido University Computing Center by parallel computation with 16
CPUs.

\end{document}